\documentclass[aps,prd,preprint,groupedaddress,showpacs]{revtex4}
\usepackage{bm,graphicx}

\begin{document}
\preprint{Rev. of: STI no. 20090423144436}
%\headline={\ifnum\pageno>1
%       \csc Semiclassical QCD Coupling \hfil
%       \folio\hfil David Batchelor \ \today
%       \else\hfil\fi}
\title{The QCD Coupling Parameter Derived \\ from the Uncertainty Principle \\
and a Model for Quark Vacuum Fluctuations}
\author{David Batchelor}
\email[]{David.A.Batchelor@nasa.gov}
\affiliation{Radiation Effects \& Analysis Group\\
National Aeronautics and Space Administration's Goddard Space Flight Center\\
Mail Code 561.4\\Greenbelt, MD  20771}
\date{July 30, 2010}
%\centerline{Short title: Semiclassical QCD Coupling}
\begin{abstract}
The magnitude of the strong interaction is characterized by $\alpha_s$,
the coupling parameter in Quantum Chromodynamics (QCD), a parameter
with an unexplained value in the Standard Model.
In this paper, a candidate explanation for $\alpha_s$ is derived
from (1) the lifetime of quark-antiquark pairs in vacuum fluctuations
given by the Uncertainty Principle,
(2) the variation of $\alpha_s$ as a function of energy 
in QCD, and (3) classical relativistic dynamics of the
quarks and antiquarks.
A semiclassical model for heavy quark-antiquark vacuum fluctuations is
described herein, based on (2) and (3).
The model in this paper predicts the measured value of
$\alpha_s(M_{Z^0})$ to be 0.121, which is in agreement with recent
measurements within statistical uncertainties.

\end{abstract}

\maketitle

\section{Introduction}

As a result of the Uncertainty Principle, vacuum fluctuations occur
consisting of a particle and its antiparticle created by the
vacuum and annihilated in a short lifetime $\Delta t$.
A fluctuation consisting of a quark and its antiquark interacting
via a gluon can occur in perturbation theory in QCD
(\cite{FGL}, \cite{GW}, \cite{HP}), the quantum field theory that
successfully describes quark interactions with great accuracy.

For educational purposes, the author developed a classical model
for the dynamics of such a quark-antiquark pair -- a model that
was intended to contrast the quantum mechanical prediction of
the pair lifetime with the classical prediction.
A surprising result was found: the pair lifetime in the classical
model agrees to a good approximation with the lifetime in quantum 
mechanics.

The model is described completely in this paper.
The model becomes semiclassical in the usual sense in a natural way.
Because the pair lifetime from the model agrees so precisely with 
the quantum mechanical lifetime, particularly in the case of bottom 
quarks, the expressions for the lifetime
in the two theories enable us to solve for the QCD strength
parameter $\alpha_s$ solely from these theoretical considerations.

In the Standard Model, QCD is a physical asymptotically free
field theory for any appropriate value of the input parameter
$\Lambda$, which is set via experimental measurement.
Within the QCD theory alone it would be impossible to establish
$\alpha_s$ without recourse to a measurement.
But the Uncertainty Principle and semiclassical QCD are not one and 
the same, and
because we find in this work that the Uncertainty Principle and 
semiclassical QCD come
to agreement on the vacuum fluctuation lifetime, we can logically
derive what the physical measurement must be.
It is shown below that this enables us to ground $\alpha_s$ on 
the value of $\hbar$.

\section{The Model Basis -- Energy Conservation}

If a quark and its antiquark are positioned at rest in their
center of mass reference frame and are released, then in general
their mutual attraction will draw them to a collision at the
origin where they will annihilate.
Photons or other particles would result, given sufficient energy.
However, the quark and antiquark experience a potential energy
$U(R)$ \cite{LSG} as a function of their separation distance
that could reduce the mass-energy of the system,
if the separation $R$ between particles is small enough.
The energy conservation relation
\begin{equation}
\varepsilon = 2\gamma\thinspace m_q c^2 + U(R) = 0 \label{Eq:energy}
\end{equation}
is possible to satisfy, where the first term is the total
relativistic energy of the particles, kinetic plus rest mass.
($R = 2r$ with $r$ the radius of either particle from the center 
of mass.)
In classical physics, the collision would not yield any energy
and so no photons or particles could be emitted.

\begin{figure}
\includegraphics[scale=0.2]{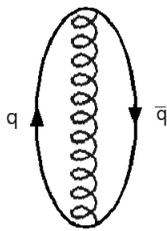}
\caption{\label{fig:vb}Feynman diagram of a virtual quark-antiquark 
pair (VQAP).
Time increases from bottom to top of figure.
The curling segment represents gluon interaction.}
\end{figure}

If the time-reversed ballistic trajectory occurred, with the
vacuum spontaneously
creating a quark-antiquark pair obeying Eq.~(\ref{Eq:energy}),
then the particles only could move apart in one-dimensional motion
to reach turning points separated by $R_{max}$.
Continuing this trajectory so that the particles fall
from the turning points back to the origin, they would
disappear back into the vacuum, like a virtual quark-antiquark pair 
(VQAP; see Fig.~\ref{fig:vb} for the Feynman diagram).
This is similar to virtual electron-positron pairs, as discussed by
Greiner (p.~3 of ref.~\cite{WG}) and Sakurai (p.~139 of ref.~\cite{JJS}).

Quantum theory implies that this two-particle system of quarks would obey 
the time-energy uncertainty relationship for the energy fluctuation
$\Delta \varepsilon$ (p.~139 of ref.~\cite{JJS})
\begin{equation}
\Delta \varepsilon \Delta t = {1\over{2}}\hbar = 5.273 \times 10^{-28}
{\rm erg\ s}. \label{Eq:teur}
\end{equation}
The energy fluctation is $\Delta \varepsilon = 2 m_q c^2$, since the
mass-energy of each quark contributes $m_q c^2$.
The quantum-mechanical lifetime of the fluctuation is $\Delta t$.
Since $\hbar$ is the quantum of action,
Eq.~(\ref{Eq:teur}) establishes an action integral {\cal A} that 
characterizes a VQAP 
that has $\Delta \varepsilon = 2 m_q c^2$.

The first purpose of this paper is to present classical
computations of $\Delta t$ and the action integral for the trajectory
described above, which turn out to give results that satisfy 
Eq.~(\ref{Eq:teur}) remarkably well, 
provided that the QCD interaction between the particles 
is well-de\-scribed by the potential energy function.

The second purpose of the paper follows from the fact that
QCD cannot specify the value of $\alpha_s$ at arbitrary energy or
4-momentum scale $Q$ 
without an established measurement of $\alpha_s$ at some particular energy 
$\mu$ \cite{SB};
but once the renormalized coupling $\alpha_s(\mu^2)$ is
measured, then QCD precisely gives the variation of $\alpha_s$ as a function
of energy (the ``running" coupling).
The present paper offers a theory based on the action integral that establishes
the value of $\alpha_s$ at the energy scale of twice the bottom quark 
mass-energy to good approximation.
This enables one to use the QCD running coupling $\alpha_s(Q^2)$
to determine the coupling strength in general in the usual way
to good approximation.

\section{QCD Potential Energy Function}

As discussed in detail by Lucha {\sl et al.}~(especially pp.~161-162
of ref.~\cite{LSG}), a potential 
energy function serves to describe the bound states of heavy quarks
(charm, bottom, and top).
For light quarks the QCD interaction
is not satisfactorily described by a potential energy function
and will not be attempted here.
Here we apply the standard potential energy treatment to the
heavy charm, bottom and top quarks.

We use the standard Cornell potential \cite{EE1}, \cite{EE2}
\begin{equation}
V(R) = - {4\over{3}}{\alpha_s\hbar\thinspace c\over{R}} + 
aR\ ,\label{Eq:cp}
\end{equation}
where $\alpha_s$ is the dimensionless QCD strong coupling strength
and $a \approx$ 0.25 GeV$^2$.
The second term, $a R$, is only significant for $R > 10^{-13}$ cm.
We will not need to consider the $a R$ term, since the first term with
the Coulomb-like dependence turns out to
strongly dominate the potential because
$R_{max} \ll 10^{-13}$ cm for VQAPs.

The model VQAP is a form of bound state.
The standard way to account for the variation of $\alpha_s(Q^2)$ 
in a quark bound state is to let $\alpha_s$
depend on the quark masses and use $Q^2 = (m_1 + m_2)^2$, 
with the $m_i$ the quark masses (see p. 129 of ref.~\cite{DBL}).
To model a VQAP we may then compute $\alpha_s$ to leading order
(Eq.~(6) of Ref.~\cite{SB}).
So the QCD coupling is then given by
\begin{equation}
\alpha_s(Q^2)
 = {1\over{\beta_0\ {\rm ln} (Q^2/\Lambda^2)}}\ ,
\label{Eq:alphas}
\end{equation}
where $\beta_0$ is defined by
\begin{equation}
\beta_0 = {33 - 2n_f\over{12\pi}}\ ,
\end{equation}
$n_f$ is the number of quark flavors with masses much less
than $m_1 + m_2$, and $\Lambda$ is the QCD scale energy, 
\begin{equation}
\Lambda^2 = {\mu^2\over{e^{1/(\beta_0 \alpha_s(\mu^2))}}}\ .
\end{equation}
$\Lambda$ is approximately 0.093 GeV, assuming that 
$\mu \equiv M_{Z^0} =$ 91.2 GeV (the mass-energy of the $Z_0$ particle),
$\alpha_s(M_{Z^0}) = 0.119 \pm 0.002$.
This is a typical value of $\Lambda$ for a one-loop approximation
(\cite{SB} p. R31).

For the running coupling near the charm quark mass $n_f = 4$, near the
bottom quark mass $n_f = 5$, and near the top quark mass $n_f = 6$.
We now can use $m_1 = m_2 = m_q$ for each quark mass $m_q$
in the expression for $Q^2$ and
compute $\alpha_s$ from Eq.~(\ref{Eq:alphas}) for a running coupling.
Then with a charm quark current mass $m_q = m_c$ of 1.27 GeV/$c^2$ we have
$\alpha_s(2m_c c^2) = 0.228$;
with a bottom quark current mass $m_q = m_b$ of 4.2 GeV/$c^2$ we have 
$\alpha_s(2m_b c^2) = 0.167$;
with a top quark current mass $m_q = m_t$ of 173.1 GeV/$c^2$ we have 
$\alpha_s(2m_t c^2) = 0.109$ \cite{PDG}.

\section{Classical ballistic trajectory and lifetime of VQAP}

Let us calculate the trajectory lifetime $t_{vq}$ for a VQAP.
If one solves the energy equation (\ref{Eq:energy}) for the dynamics using 
$V(R)$, the nonrelativistic potential energy, then the particle velocities 
nevertheless exhibit relativistic motion, approaching $c$ 
asymptotically at the origin $r = 0$.
Thus it is necessary to correct the potential energy for 
relativistic effects.
Jackson (\cite{JDJ}, p.~553) demonstrates how this is done by using
the relativity factor $\gamma$ defined in Eq.~(\ref{Eq:gamma}) below.
For this trajectory of linear motion, the transformation 
$R \rightarrow \gamma\thinspace R$ in the expression for $V(R)$ performs
the appropriate modification of the potential energy function 
(since we are considering the center-of-mass reference frame).
The potential energy function in Eq.~(\ref{Eq:cp}) becomes
\begin{equation}
U(R) = - {4\over{3}}{\alpha_s\over{\gamma R}}\ \hbar \thinspace c.
\label{Eq:u}
\end{equation}
With this $U(R)$ we can solve Eq.~(\ref{Eq:energy}) for $R_{max}$ at the 
turning point (where $\gamma = 1$):
\begin{equation}
2\thinspace m_q\thinspace c^2 \equiv -\thinspace U(R_{max})
 = {4\over{3}}{\alpha_s\over{R_{max}}}\hbar \thinspace c
\end{equation}
\begin{equation}
R_{max} = {2\over{3}}{\alpha_s \hbar \over{m_q\thinspace c}}\ .
\label{Eq:rmax}
\end{equation}
For the charm quark current mass $m_c$ of 1.27 GeV/$c^2 \equiv 2.26 
\times 10^{-24}$ g, we find the charm VQAP has an $R_{max} =
2.35 \times 10^{-15}$ cm.

Checking the terms in Eq.~(\ref{Eq:cp}) for $V(R_{max})$ shows that the 
first term is about -100 times the second term.
This confirms that the quarks are so deep in the potential well
that the $a R$ term of the Cornell potential
can be neglected in solving the problem.

Let us define the time from appearance of a quark at the origin
$r = 0$ to the time that the quark stops at the turning point
$r = {1\over 2}R_{max}$ as ${1\over 2}t_{vq}$.
The $t_{vq}$ is the classical equivalent of the quantum-mechanical
$\Delta t$ that we seek.
We note that
\begin{equation}
\gamma^2 \equiv {1\over{1 - \beta^2}}\qquad 
\beta = {1\over{c}}{dr\over{dt}} \label{Eq:gamma}
\end{equation}
and solve for $dt$, which we shall integrate.
We rewrite the energy equation (\ref{Eq:energy}) with 
$\zeta \equiv R/R_{max}$ as 
\begin{equation}
\gamma^2 = \zeta^{-1}\ \Rightarrow dt
 = {dr\over{c\sqrt{1-\zeta}}}\ .
\end{equation}
The time for the particle to fall from $r = R_{max}/2$ back to $r = 0$
is also ${1\over 2}t_{vq}$, so we have
\begin{equation}
t_{vq} = 2\int_0^{R_{max}/2}{dr\over{c\sqrt{1 - \zeta}}}
 = {R_{max}\over{c}}
 \int_0^1 {{d\zeta}
           \over{ \sqrt{1 - \zeta }}  }
 = {R_{max}\over{c}}
    {\sqrt{\pi}\thinspace \Gamma(1)\over{\Gamma({3\over{2}})}}
 = {2 R_{max}\over{c}}
 = {4\over{3}}{\alpha_s\hbar\over{m_q c^2}} \label{Eq:tvq}
\end{equation}
The integral is given in ref.~(\cite{GAKTMK}, p.~974).
The value of $t_{vq}$ is the total time for either quark 
to travel from $r = 0$ to its turning point and back to $r = 0$.

For the charm quark, with $m_q = m_c \approx
2.26 \times 10^{-24}$ g, we find the trajectory lifetime of the
VQAP to be $t_{vq} \approx 1.57 \times 10^{-25}$ s.
In comparison, the standard lifetime of the charm VQAP, given by 
the Uncertainty Principle expressed in Eq.~(\ref{Eq:teur}), is
$\Delta t = \hbar/(4 m_c c^2) \approx 1.29 \times 10^{-25}$ s.
So $t_{vq}$ from the classical computation is approximately
22\% larger than $\Delta t$.
This is remarkably close agreement of the classical lifetime
with the quantum mechanical lifetime.

For the bottom quark, with $m_q = m_b \approx
7.48 \times 10^{-24}$ g, we find the trajectory lifetime of the
VQAP to be $t_{vq} \approx 3.47 \times 10^{-26}$ s.
In comparison, the standard lifetime of the bottom VQAP, given by 
the Uncertainty Principle expressed in Eq.~(\ref{Eq:teur}), is
$\Delta t = \hbar/(4m_b c^2) \approx 3.90 \times 10^{-26}$ s.
So $t_{vq}$ from the classical computation is approximately
11\% smaller than $\Delta t$.
This also is remarkably close agreement of the classical lifetime
with the quantum mechanical lifetime.

For the top quark, with $m_q = m_t \approx
3.08 \times 10^{-22}$ g, we find the trajectory lifetime of the
VQAP to be $t_{vq} \approx 5.51 \times 10^{-28}$ s.
In comparison, the standard lifetime of the top VQAP, given by 
the Uncertainty Principle expressed in Eq.~(\ref{Eq:teur}), is
$\Delta t = \hbar/(4m_t c^2) \approx 9.48 \times 10^{-28}$ s.
So $t_{vq}$ from the classical computation is approximately
42\% smaller than $\Delta t$.
This also is remarkably close to the quantum mechanical lifetime.

\section{Action integral for the trajectory}

A key step in quantizing a classical model to make it a
semiclassical model of a quantum system is computation of
the action integral.
In the present model, that is done as follows.
The expression for the integral of action associated with
a potential function $U$ acting on a particle, in the
relativistic case, is given by Lanczos (\cite{CL}, p. 321):
\begin{equation}
{\cal A} = - \int_{t_1}^{t_2} U {ds\over{c}}
\end{equation}
where $ds = c\thinspace dt/\gamma$.
Considering the integrated action of the potential energy 
field in a VQAP, we compute the field action integrated
over $t_{vq}$:
\begin{equation}
{\cal A} =
 - 2 \int_0^{t_{vq}/2} \biggl(-{4\alpha_s \hbar c
 \over{3 \gamma R}}\biggr) {dt\over{\gamma}}
 = {8\alpha_s\hbar\over{3}} \int_0^{R_{max}/2} {dr\over{
                               \gamma^2 R \sqrt{1 - \zeta}}}
\end{equation}
\begin{equation}
= {4\alpha_s\hbar\over{3}}
 \int_0^1 {d\zeta\over{\zeta^{-1}\zeta \sqrt{1 - \zeta}}}
 = {4\alpha_s\hbar\over{3}}
 \int_0^1 {d\zeta\over{\sqrt{1 - \zeta}}}
 = {8\alpha_s\over{3}}\hbar\label{Eq:action}
\end{equation}
For the charm VQAP, $\alpha_s(2m_c c^2) = 0.228$ and therefore
${\cal A} = 0.61\thinspace \hbar$.
This action integral is only 22\% larger than the exact VQAP 
quantum fluctuation action in Eq.~(\ref{Eq:teur}), ${1\over{2}}\hbar$.

In the case of the bottom quark,
the action integral for the model of the VQAP is found by
substituting $\alpha_s(2m_b c^2) = 0.167$ into Eq.~(\ref{Eq:action}), 
and we find 
${\cal A} = 0.45\thinspace \hbar$.
This is 10\% lower than the quantum mechanical action for a VQAP.

In the case of the top quark,
the action integral for the model of the VQAP is found by
substituting $\alpha_s(2m_t c^2) = 0.109$ into Eq.~(\ref{Eq:action}), 
and we find 
${\cal A} = 0.29\thinspace \hbar$.
This is 42\% lower than the quantum mechanical action for a VQAP.

\section{Spin-spin Interaction Effects}

As noted by Lichtenberg (p.~133 of ref.~\cite{DBL}) the spin-spin interaction
between quarks is the source of electromagnetic mass splittings among
hadron isospin multiplets.
Here it will be shown that spin-spin interaction between the quark and
antiquark in a VQAP is significant for the charm and top quark cases.
Spin-spin interaction would modify the dynamics of the semiclassical model
and lead to a different action integral and trajectory lifetime, so its
influence needs to be quantified.

The creation of a quark-antiquark pair from the vacuum should conserve
angular momentum, so we consider the case in which the particles have
spin parallel to the motion axis but pointing in opposite directions.
The potential energy function in Eq.~(\ref{Eq:u}) becomes (\cite{JDJ} p.~185)
\begin{equation}
U(R) = - {4 \alpha_s \over{3 \gamma R}}\hbar c
       - {2 \mu_q^2\over{(\gamma R)^3}}
\end{equation}
where
\begin{equation}
\mu_q = {gQ_q\over{2 m_q c}}{\hbar\over{2}}
\end{equation}
is the magnetic moment of a quark with electric charge $Q_q$ 
and $g$ is the Land\'e factor,
which will be taken as equal to 2 for present purposes.

The energy equation Eq.~(\ref{Eq:energy}) in this case becomes 
\begin{equation}
\varepsilon = 2\gamma\thinspace m_q c^2 
- {4\over{3}}{\alpha_s\over{\gamma R}}\ \hbar \thinspace c
- {2\mu_q^2\over{(\gamma R)^3}} = 0 \label{Eq:energy2}
\end{equation}
Again we can derive the separation of the particles at the
turning point $R_{max}^\prime$ by letting $\gamma \rightarrow 1$.
It is convenient to use $\zeta = R/R_{max}$ as a dimensionless variable
again, and with that substitution, the energy equation becomes
\begin{equation}
\zeta_{max}^3 - \zeta_{max}^2 - \Delta = 0 \qquad {\rm with} 
\qquad \Delta = {27 Q_q^2\over{32 \alpha_s^3\hbar c}}\ .
\end{equation}
The parameter $\Delta$ is a measure of the magnitude of the
spin-spin interaction relative to the QCD interaction;
$\Delta \rightarrow 0$ recovers the previous case in Eq.~(\ref{Eq:energy}).

Solutions of this new energy equation (\ref{Eq:energy2}) are deferred for a later
paper, but the values of $\Delta$ for each of the heavy quarks
are illuminating.
The electric charges of the charm and top quarks are both
${2\over 3} e$, but the charge of the bottom quark is only
$-{1\over 3} e$ .
Consequently the parameter $\Delta$ is 0.23 for the charm quark
and 2.1 for the top quark.
For the bottom quark, $\Delta = 0.15$, which indicates that the
importance of spin-spin effects in the dynamics of the model
is minimal for the bottom quark.

The smallest value of $\Delta$ is associated with the quark
that exhibits the best agreement between $\Delta t$ and $t_{vq}$,
the bottom quark.
This suggests that the bottom quark case is best for using the
model to characterize the QCD interaction without perturbations
from the spin-spin effects.

In the above cases,
the model's representation of the bottom VQAP inherently 
is approximately quantized -- a remarkable agreement between a
quantum-mechanical characteristic of a dynamical system and
the classical description of it.
In comparison, semiclassical models for mesons, which achieve
excellent agreement with measurements of meson masses \cite{FB}, 
\cite{SV} need to be formulated with additional
quantization conditions that introduce the factor $\hbar$.
We have not imposed any quantization conditions upon the trajectory
in this dynamical model.
The bottom quark model herein achieves approximate quantization at 
$\alpha_s(2m_b c^2)$
based upon only the measured value of $\alpha_s(M_{Z^0})$, the QCD
theoretical energy dependence of $\alpha_s(Q^2)$, and 
relativistic dynamical theory (Eqs.~(\ref{Eq:energy}) and 
(\ref{Eq:u})).

Preliminary work to investigate the discrepancy between $t_{vq}$ 
and $\Delta t$ in the cases of the charm quark
and top quark has been performed as the author will show 
elsewhere \cite{DB}:
for the charm quark and top quark, electromagnetic spin-spin 
interactions become
important and influence $t_{vq}$ and ${\cal A}$ in such a way
as to bring into closer agreement the classical and quantum results.

\section{Conclusions}

The salient logic in this paper's result is the following.
The measurement of $\alpha_s(M_{Z^0})$ and the one-loop $\Lambda$
obtained from the so-called `modified minimal subtraction scheme'
of renormalization theory \cite{SB} predict $\alpha_s(2 m_b c^2)$.
From this we may use the classical ballistic trajectory lifetime 
of the VQAP to compute the $t_{vq}$ and action $\cal A$, obtaining 
${\cal A} \approx 
{1\over 2}\hbar$ in approximate agreement with quantum mechanics.
From this we may go further.

Since $\hbar$ is a more universal and fundamental parameter than
$\alpha_s$, $\hbar$ intuitively would seem to be the governing
parameter in the action equation Eq.~(\ref{Eq:action}).

If $\alpha_s(2 m_b c^2)$ equalled 3/16 then ${\cal A}$ would exactly
equal ${1\over 2}\hbar$.
Setting Eq.~(\ref{Eq:alphas}) equal to 3/16, $Q^2$ equal to $(2m_b c^2)^2$, 
and solving for 
$\Lambda$ yields $\Lambda = 0.106$ GeV instead of the standard 
0.093 GeV.
With this value of $\Lambda$, Eq.~(\ref{Eq:alphas}) gives 
$\alpha_s$(91.2 GeV $\equiv M_{Z^0}$) = 0.121.
This is only 2\% different from the measured value upon which
the accuracy of QCD depends, and is within the statistical
uncertainty in $\alpha_s(M_{Z^0})$ quoted in Ref.~\cite{SB}.

We now have a mathematical link between the measured $\alpha_s(M_{Z^0})$
and the action integral for QVAPs from the Uncertainty Principle,
${1\over 2}\hbar$.
This logical sequence is equally valid in reverse.
We may take as starting point the action ${\cal A}$ and infer that 
the action integral ${1\over 2}\hbar$ is what governs the
value of $\alpha_s(M_{Z^0})$.
This reverse argument from $\cal A$ = ${1\over 2}\hbar$ through 
Eqs.~(\ref{Eq:action}), (\ref{Eq:tvq}), and (\ref{Eq:alphas}) to 
$\alpha_s(M_{Z^0})$
is the derivation mentioned in this paper's title.

The Uncertainty Principle is more general than QCD, and the
reason that this derivation works is that the fluctuation
lifetime $\Delta t$ is a different condition than the semiclassical
model lifetime which is derived from phenomenological QCD.
Because these two theoretical lifetimes agree in the special
bottom quark case, we can now consider the QCD input parameter
$\alpha_s(M_{Z^0})$ to be derivable from our theories.

This good agreement between the classical trajectory lifetime and
the quantum uncertainty lifetime at the key mass-energy of the
bottom quark is surprising, but it may have a
simple physical explanation: if the vacuum creates these
particles in motion at $v \approx c$, then their de Broglie 
wavelengths $\lambda = h/p$ should be small, and wave packets that
are small relative to $R_{max}$ would represent the particles well.
Ballistic dynamics of point masses then would serve as a good
approximation for the particle motions and the
agreement of the ballistic and quantum mechanical timescales would
be accounted for.
The length scale of any VAP is usually
characterized in standard literature by assuming that 
$v \approx c$ \cite{JJS}.

The semiclassical model described herein may be used to produce
predictions for experiments involving VQAPs that are interacting
with other particles, instead of the unobserved VQAPs that were 
modelled in this paper.

\begin{acknowledgments}
% put your acknowledgments here.
The author is grateful to NASA's Goddard Space Flight Center for 
support during this research.
\end{acknowledgments}

%\bibliography{}

\end{document}